\documentclass[a4paper,11pt]{article}
\pdfoutput=1
\usepackage{amssymb}
\usepackage{textcomp}
\usepackage[utf8]{inputenc}
\usepackage{latexsym}
\usepackage{epstopdf}
\usepackage{epsfig}
\usepackage{graphicx}
\usepackage{amsmath}
\usepackage{multirow}
\usepackage{subfigure}
\usepackage{a4wide}
\usepackage{cite}
\usepackage{rotating}
\usepackage{hyperref}
\newcommand{\be}{\begin{equation}}
\newcommand{\ee}{\end{equation}}
\newcommand{\ba}{\begin{eqnarray}}
\newcommand{\ea}{\end{eqnarray}}
\newcommand{\bac}{\begin{array}{c}}
\newcommand{\eaa}{\end{array}}
\newcommand{\baz}{\begin{array}{cc}}
\newcommand{\mathsym}[1]{{}}

\newcommand{\bad}{\begin{array}{ccc}}

\newcommand{\bi}{\begin{itemize}}
\newcommand{\ei}{\end{itemize}}
\newcommand{\bmt}{\begin{pmatrix}}
\newcommand{\emt}{\end{pmatrix}}
\newcommand{\bt}{\begin{tabular}}
\newcommand{\et}{\end{tabular}}

\newcommand{\ovl}{\overline}

\newcommand{\benu}{\begin{enumerate}}
\newcommand{\eenu}{\end{enumerate}}

\def\dsodt{ds_{23}\over dt}
\def\dstdt{ds_{13}\over dt}
\def\dsthdt{ds_{12}\over dt}

\def\be{\begin{equation}}
\def\ee{\end{equation}}
\def\ba{\begin{eqnarray}}
\def\ea{\end{eqnarray}}
\def\br{\begin{array}}
\def\er{\end{array}}

\def\ol{\overline}

\def\op{\oplus}

\def\Sg0{\Sigma_0}
\def\Sg1{\Sigma_1}
\def\Sg2{\Sigma_2}

\def\Dl{\Delta_L}

\def\Dr{\Delta_R}
   
\def\al{\alpha}
\def\al0{\alpha_0}
\def\al1{\alpha_1}
\def\al2{\alpha_2}
\def\al3{\alpha_3}
\def\al4{\alpha_4}
\def\al5{\alpha_5}
\def\bt0{\beta_0}
\def\bt1{\beta_1}
\def\bt2{\beta_2}
\def\bt3{\beta_3}
\def\bt4{\beta_4}
\def\bt5{\beta_5}
\def\gm0{\gamma_0}
\def\gm1{\gamma_1}
\def\gm2{\gamma_2}
\def\dl0{\delta_0}
\def\dl1{\delta_1}
\def\dl2{\delta_2}
\begin{document}
\title{\Large\bf{ Unification Predictions With or Without Supersymmetry }}

\author{\bf M.K. Parida$^{*}$, Riyanka Samantaray$^{\dagger}$ \\
Centre of Excellence in Theoretical and Mathematical Sciences, \\
Siksha  O Anusandhan (Deemed to be University), Khandagiri Square,\\
 Bhubaneswar 751030, Odisha,  India}

\maketitle

\begin{abstract}
Supersymmetric (SUSY) grand unified theories (GUTs) appear to be best
motivated for understading   strong, weak and electromagnetic
interactions of nature. We briefly review emergence of new formulas for
running fermion masses valid in direct breaking of GUTs. High scale
mixing unification of quark and neutrino mixings and  existence of
theorems on vanishing theoretical uncertainties in GUT predictions are
discussed.  D-Parity properties of SO(10)
representations leading to large number of intermediate breaking
models are pointed out. Unification predictions of SUSY SO(10) in the light of neutrino mass, lepton flavor violation, 
baryogenesis via leptogenesis within gravitino constraint, and proton
decay are noted. We further
discuss realisation of flavour unification and possibility of fitting all
fermion masses through  R-Parity and D-Parity conserving left-right symmetric 
intermediate breaking in SUSY SO(10)$\times S_4$. In the absence of SUSY, 
 two interesting possibilities of minimal  grand desert modifications by
  only one intermediate mass scalar  in each case and their
 applications to  dark
 matter decay through type-I seesaw are briefly noted. Heavy scalar
 triplet decay  leptogenesis through new ansatz for type-II seesaw dominance in non-SUSY SO(10), emergence of new CP
 asymmetry formulas and model capabilitities  to explain WIMP dark matter, vacuum stability of the scalar potential and experimentally observed limit on proton lifetime are briefly summarised.
\end{abstract}
\noindent{${}^*$email:minaparida@soa.ac.in}
\section{Introduction}\label{sec:intr}
The standard model of electroweak
and strong interaction gauge theory, $SU(2)_L \times U(1)_Y\times
SU(3)_C$, enjoys a very special staus in the fundamental understanding
of particle interaction and three forces of nature. It was a much
saught after theoretical breakthough after Dirac theory of Quantum
Electrodynamics (QED)  \cite{Dirac:1931} which achieved precision
prediction in higher orders of electromagnetic gauge coupling 
successfully through its intrinsic capability of
renormalizability. Dirac's idea manifested in the generalization
Yang-Mills Lagrangian for non-Abelian gauge theories \cite{Yang-Mills}. In sharp contrast with
massless photon of $U(1)_Q$ invariant QED, a major hurdle in achieving a
renormalizable SM was the compelling experimental and phenomenological
issues that demanded massive vector bosons to mediate weak
interaction. The ingenuous idea of gauging the electroweak theory
\cite{Glashow:1962,Weinberg:1967,Salam:1968} combined with Higgs
mechanism \cite {Higgs,Higgsexpt:2012}
finally resolved the long standing issue with the emergence of renormalisable
 electroweak theory even after its spontaneous symmetry breaking \cite{Veltman}. \\ 
Even though the SM has been tested by numerous experiments , it fails to 
explain several issues, the most prominent being neutrino oscillation \cite{nudata,Forero:2014bxa,Esteban:2018azc}, baryon asymmetry of
the universe (BAU) \cite{BAUexpt,Planck15}, nature of dark matter (DM)
and its stability \cite{DMexpt,GAMBIT}, and the origin of disparate values of gauge couplings. Besides these the SM faces the most fundamental issue of protecting the Higgs mass at the electroweak scale. This is due to the fact that the SM Higgs mass
becomes qudratically divergent by radiative corrections against which there 
does not seem to be any natural solution except through supersymmetery
\cite{SUSYrev1,SUSYrev2,Nath:2007,Dimo-Raby-Wil:1981,Witten:1981,Dimo-Georgi,Kaul:1982,Marciano-gs:1982,Amaldi:1991}.       
 Grand unified theories (GUTs) \cite{ps:1974,Pati:2008,su5,so10,E6},
 originally aimed at unifying the three forces of nature, were
 subsequently supersymmetrised to confront the gauge hierarchy
 problem, exhibit explicit coupling unification through direct
 breaking to SM and address issues like neutrino masses and WIMP dark matter. 

Including Fermi-Bose symmetry, the particle content of  minimal
supersymmetric standard model (MSSM) is shown in Table \ref{tab:mssmp}. The second column of
Table \ref{tab:mssmp} represents the SM particle content of one fermion generation except that instead of two Higgs dublets of MSSM, SM has only the standard doublet $\phi (2,1/2,1)$. The underlying Fermi-Bose symmetry of MSSM and SUSY GUTs naturally
cancels out the quadratic divergence of the Higgs mass thus removing the gauge hirarchy problem.       
Another major theoretical achievement of MSSM descending from SUSY
GUTs is the automatic natural explanation of three forces of SM as
discussed below.
\section{Renormalization Group Evolution of Couplings and Masses}
\subsection{SUSY Grand Desert Unification}
To understand failure of unification in SM and its success in MSSM and
SUSY GUTs, the precision
electroweak measurements are used to detemine the SM gauge couplings
at the electroweak scale
\be
{\alpha_{Y}}^{-1}(M_Z)= 59.8,\,{\alpha_{2L}}^{-1}(M_Z)=29.6,\,
{\alpha_{3C}}^{-1}(M_Z)=8.54, \label{eq:invai}
\ee
where $\alpha_i=g_i^2/(4pi)$.
The evolution of gauge couplings are given by the renormalisation
group equations (RGEs) \cite{GQW,Jones:1982,Langacker:1994}
\be
\mu\frac{{\partial g}_i}{{\partial
   \mu}}=\frac{a_i}{16\pi^2}+\frac{g_i^3}{(16\pi^2)^2}\left(\sum_j b_{ij}g_j^2
-k_iy_{top}^2\right), \label{eq:rge} 
\ee 
where $a_i(b_{ij})$ are one-loop (two-loop) coefficients, and
$k_i=(17/10,3/2,2)$ for SM but $k_i=(26/5,6,4)$ for MSSM.
Denoting the Dynkin indices due to gauge bosons,
fermions, and Higgs scalars by $t_2(G_i), t_2(F_i)$ and
$t_2(S_i)$, respectively, under gauge group $G_i$ the analytic
formulas for
  one-loop beta function coefficients in non-supersymmetric (non-SUSY) and SUSY cases are.\\
\par\noindent{\bf Non-SUSY Gauge Theory:}\\ 
\ba
a_i&=&-\frac{11}{3}t_2(G_i)+\frac{2}{3}t_2{(F_i)}+\frac{1}{3}t_2{(S_i)},\,\, (i\in SU(N)),
\nonumber\\
 &=& \frac{2}{3}t_2{(F_i)}+\frac{1}{3}t_2{(S_i)},\,\,(i\in U(1)), \label{eq:nonsusyai}
\ea
\par
\begin{figure}[h!]
\includegraphics[scale=0.25]{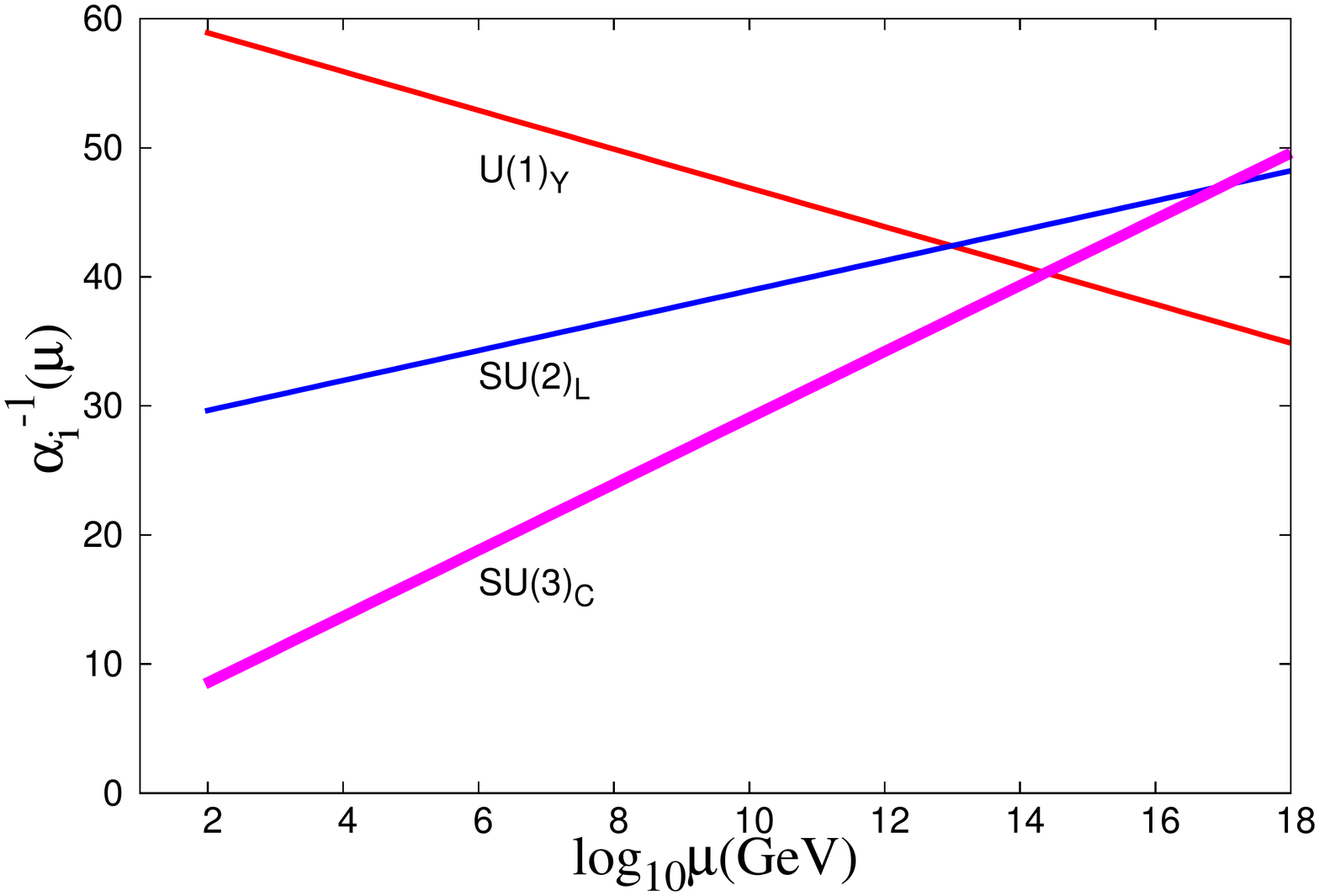}
\includegraphics[scale=0.25]{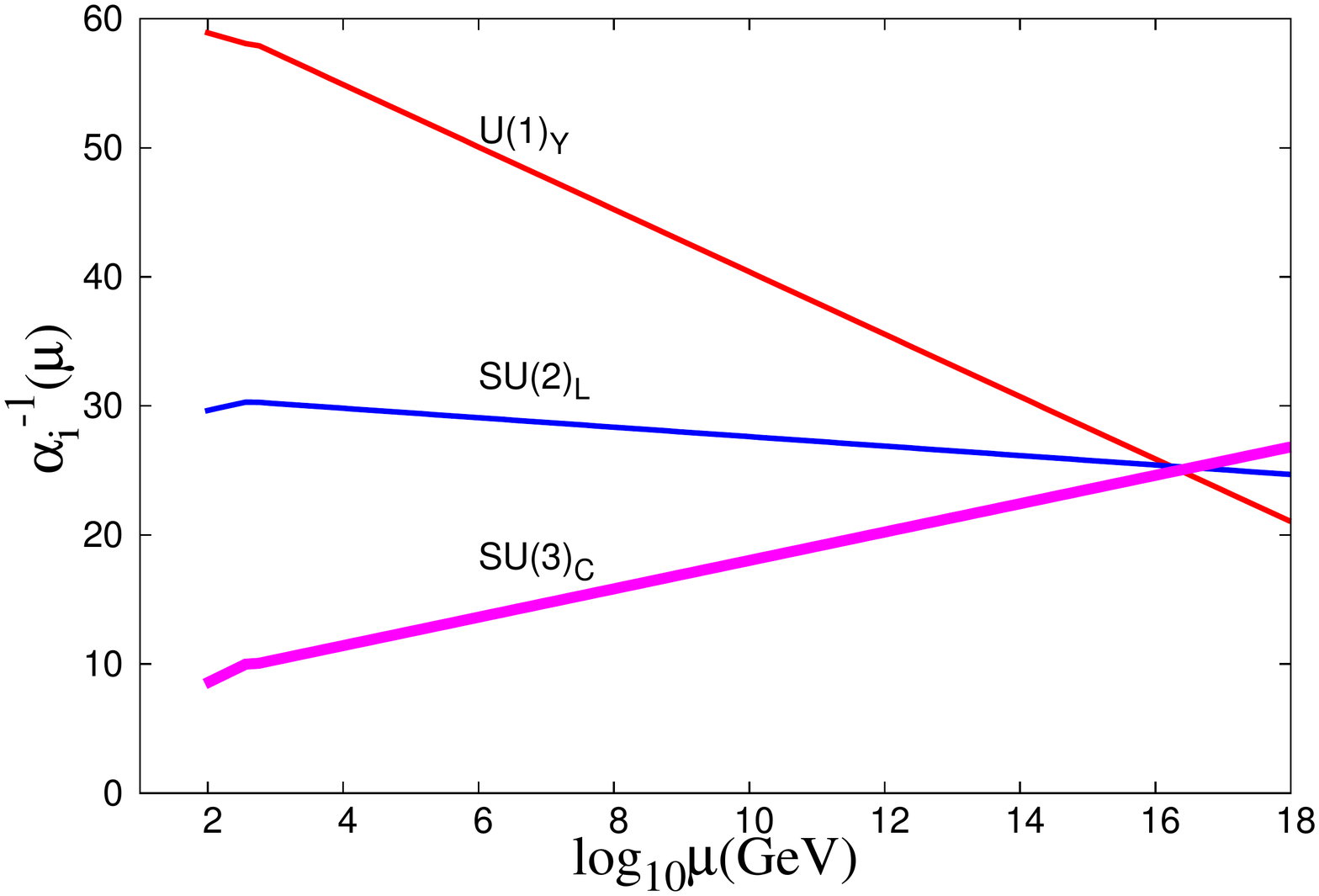}
\caption{ Evolution of  the three
  inverse fine strucure constants without unification  in the
SM (Left panel), but with  unification in the MSSM (Right panel).}
\label{fig:mssmu}
\end{figure}

\noindent{\bf SUSY Gauge Theory:}\\
\ba
a_i&=&-\left[\frac{11}{3}t_2(G_i)+\frac{2}{3}t_2{(G_i)}\right] 
+\left[\frac{2}{3}t_2{(F_i)}+ \frac{1}{3}t_2{(F_i)}\right]
\nonumber\\
&+&\left[\frac{1}{3}t_2{(S_i)}+\frac{2}{3}t_2{(S_i)}\right] \nonumber\\
&=&-{3t_2(G_i)}+t_2{(F_i)}+t_2{(S_i)},\,\, (i\in SU(N)),
\nonumber\\
 &=& t_2{(F_i)}+t_2{(S_i)},\,\,(i\in U(1)), \label{eq:aisusy}
\ea
The first two lines in eq.(\ref{eq:aisusy})  is derived from Non-SUSY eq.(\ref{eq:nonsusyai}) using Fermi-Bose symmetry.
Similar analytic formulas exist for two-loop coefficients $b_{ij}$ of
eq.(\ref{eq:intrge}) given below.
Then $a_i=(41/10,-19/6,-7)$ for SM but $a_i=(33/5,1,-3)$ for MSSM for which particle contents are shown in   Table \ref{tab:mssmp}. 
\begin{table*}
\begin{center}
\begin{tabular}{|c|c|c|}
\hline
Particle Type & $G_{213}$ Charges & Superpartners $\&$ Charges   \\
\hline
&&\\
 ${\rm Gauge\, Bosons}$ & $W_\mu(3,0,1),B_{\mu}(1,0,1), G_{\mu}(1,0,8)$ & ${\tilde W}_{\mu}(3,0,1),{\tilde B}_{\mu}(1,0,1),{\tilde G}_{\mu}(1,0,8)$\\
&&\\
\hline
&&\\
${\rm Fermions\,of\,ith }$ & $l_i(2,-1/2,1),e_{R_i}(1,-1,1)$ & ${\tilde l}_i(2,-1/2,1),{\tilde e}_{R_i}(1,-1,1) $ \\
 generation&&\\
&$Q_i(2,1/6,3),u_{R_i}(1,2/3,3), d_{R_i}(1,-1/3,3)$&${\tilde Q}_i(2,1/6,3),{\tilde u}_{R_i}(1,2/ 3,3),{\tilde d}_{R_i}(1,-1/3,3)$\\
&&\\
\hline
&&\\
${\rm Higgs\,Scalars}$ & $\phi_u(2, 1/2, 1), \phi_d(2,-1/2, 1)$& ${\tilde \phi}_u(2, 1/2, 1), {\tilde \phi}_d(2,-1/2, 1)$ \\
&&\\
\hline
\end{tabular}
\end{center}
\caption{Particle content of MSSM for three generations of fermions ($i=1,2,3$).
The SM partcle content is devoid of superpartners and has only one Higgs scalar doublet $\phi (2,1/2,1)$ instead of two.}
\label{tab:mssmp}
\end{table*}
These coefficients are used in the integral form of evolution equations 
\be
\frac{1}{\alpha_i}(\mu)=\frac{1}{\alpha_i}(M_Z)-\frac{a_i}{2\pi}\ln(\frac{\mu}{
M_Z})
-\frac{1}{4\pi}\sum_jB_{ij}\ln[\frac{\alpha_j(\mu)}{\alpha_j(M_Z)}], 
\label{eq:intrge}
\ee
where $\alpha_i(\mu)=\frac{g_i^2(\mu)}{4\pi}$ and
$B_{ij}=b_{ij}/a_j$.  The second (third) term in the RHS represent one
(two-loop) effect \cite{Jones:1982}. 
For simplicity, the evolution  of  gauge couplings at one-loop level is shown
 in Fig.\ref{fig:mssmu} where the left (right) is for SM (MSSM).

The presence of a triangle of finite area in the case of SM, instead of a single meeting
point (or a much smaller triangle compatible with experimental errors),

demonstrates inherent deficiency of the minimal SM to unify the three
gauge couplings. On the other hand profound unification is exhibited in
MSSM with the unification scale $2\times 10^{16}$
GeV.\cite{Amaldi:1991}. Thicker sizes of the three curves in SUSY
case arises due to existing uncertainty at the electroweak scale.

About eight years before CERN-LEP data inspired SUSY unification was noted \cite{Amaldi:1991}, such unification was also observed in non-SUSY SO(10) GUT  with left-right 
intermediate gauge symmetries like $SU(2)_L\times SU(2)_R\times
U(1)_{B-L}\times SU(3)_C$($\equiv G_{2213}$), $SU(2)_L\times
SU(2)_R\times SU(4)_C$ ($\equiv G_{224}$), $G_{2213D}$ and $G_{224D}$
  where symmetries with $D$ stand for D-Parity indicating
  $g_{2L}=g_{2R}$ \cite{cmp-PRL:1984}. It was further noted that each of the non-SUSY
  GUTs like SO(10) or $E_6$ can accommodate one or more intermediate
  gauge symmetry breaking providing a variety of models for non-SUSY
  grand unification \cite{cmp-PRD:1984} that has resulted in a number of
  interesting applications \cite{cmgmp:1985} including prospects of low mass $W_R,Z_R$
  bosons.    

\subsection{New Formulas for Running Fermion  Masses in SM, 2HDM, and MSSM }\label{sec:run}
Defining $\tan\beta=v_u/v_d=$ where $v_u(v_d)=\langle \phi_u(\phi_d)\rangle$, 
a very attractive aspect of MSSM and SUSY GUTs is $b-\tau$ Yukawa
unification for smaller  $\tan{\beta}\sim 1-5$  and
approximate $t-b-\tau$ Yukawa unification for larger
$\tan{\beta}\sim 40-55$. SUSY
SO(10) besides gauge and Yukawa unification, also
possesses ability for reasonable parametrisation of charged fermion masses
at the GUT scale \cite{Babu-RNM:1993}. The new
formulas for running fermion masses \cite{dp:2001}  were derived  taking into accoune the scale dependence of VEVs $v_u,V_d$ in MSSM and 2HDM.
\cite{Arason:1992,BLP:1993,Chankowski:1993} which were also found to
decrease with increasing mass scale instead of remaining
constant. Using the correspoding running VEVS in MSSM, SM and 2HDM  new formulas have been developed to
extrapolate all charged fermion masses from their low energy values to
the GUT scale values \cite{dp:2001}. These extrapolated values  have been found  useful
in testing SO(10) model capabilities for representing fermion masses  even without using flavour symmetries. 

\section{Unification of Quark and Neutrino Mixings}\label{sec:humt}
\subsection{Radiative Magnification with Quasi-Degenerate Neutrinos}
In the presence of supersymmetry it was found that the mixing angle
between two light neutrinos could be quite small near the SUSY GUT
scale. But due to renrmalisation group evolution, the mixing angle
gets magnified to be compatible with its large value in concordance
with neutrino oscillation data. Initially this was realised
only for atmospheric neutrino mixings. Radiative magnification
was noted to be possile for two neutrinos with (i) quasidegenerate neutrino masses, (ii)
identical CP properties, and (iii) larger
values of $\tan \beta=v_u/v_d$  \cite{Balaji-PRL:2000}.
\subsection{High Scale Mixing Unification}\label{sec:humsub}
Despite the radiative magnification mechanism that applied for the
atmospheric neutrino mixing, it was difficult to reconcile with
neutrino data showing the general behaviour of three
large values of neutrino mixings ($\theta_{ij}$)  compared to correspondingly small
quark mixings ($\theta_{ij}^q$) :
$\theta_{23}>> \theta_{23}^q, \theta_{12} >> \theta_{12}^q,
\theta_{13}>>\theta_{13}^q$. In addition the initial input value of
the mixing angle, dynamical origin of
quasi-degenerate neutrinos and the necessity of large value of
$\tau$-Yukawa coupling in the RGE were used as a matter of necessity \cite{Balaji-PRL:2000,Balaji-PLB:2000,Balaji-PRD:2001} without deeper
theoretical understanding. Through further development of neutrino RGEs \cite{mpr-PASCOS:2003,mpr-PRD:2003} an interesting resolution of this puzzle has been
suggested \cite{mpr-PRD:2003} using the underlying quark lepton symmetrty of
supersymmetric Pati-Salam theory \cite{ps:1974} (or SUSY SO(10)) along
with $S_4$ flavour symmetry. The $G_{224D}\times S_4$ breaking through
RH triplet VEV ($\langle \Delta_R \rangle \simeq V_R\simeq M_{GUT}$)
generated small departure from degeneracy created through type-II
seesaw induced
VEV of LH triplet ($\Delta_L(3,1,10)$) at the highest scale. Because
of Pati-Salam symmetric quark-lepton unification, identification of
initial boundary values of quark mixings with lepton mixings was a natural pre-existing
input for the RG evolutions. Large $\tau$-Yukawa coupling and large
value of $\tan\beta\simeq 40-55$ was a necessary prediction of
$b-\tau$ unification as shown earlier \cite{dp:2001}.
   In this theory neutrino mixings are
predicted to be unified with corresponding quark mixings at the SUSY
GUT scale. The RGEs predict negligible changes for quark mixings because of their
 strong mass hirachy. On the other hand for large $\tan\beta$ and due to quasi-degenerate
 masses, $m_{\nu_1}\simeq m_{\nu_2}\simeq
m_{\nu_3}\ge 0.2 $ eV, the RGEs for neutrino mixings are magnified to
their large low-energy values \cite{mpr-PRD:2003,mpr-others,ampr:2007,mpr-split:2005}.   
The RGEs for the mass eigen values  can be written  in a simpler form \cite{mpr-PRD:2003,mpr-others,ampr:2007}
\par
\noindent
\be
{dm_i\over dt}=-2F_{\tau}m_iU_{\tau
i}^2-m_iF_u,\,\left(i=1,2,3\right).\label{eq2}
\ee
\par
\noindent
For every $\sin\theta_{ij}=s_{ij}$, the corresponding RGEs are,
\par
\noindent
\ba
\dsodt&=&-F_{\tau}{c_{23}}^2\left( 
-s_{12}U_{\tau1}D_{31}+c_{12}U_{\tau2}D_{32}
\right),\label{eq3}\\
\dstdt&=&-F_{\tau}c_{23}{c_{13}}^2\left( 
c_{12}U_{\tau1}D_{31}+s_{12}U_{\tau2}D_{32}
\right),\label{eq4}\\
\dsthdt&=&-F_{\tau}c_{12}\left(c_{23}s_{13}s_{12}U_{\tau1}
D_{31}-c_{23}s_{13}c_{12}U_{\tau2}D_{32}\right.\nonumber \\
&&\left.+U_{\tau1}U_{\tau2}D_{21}\right).\label{eq5}
\ea
\par
\noindent
where $D_{ij}={\left(m_i+m_j)\right)/\left(m_i-m_j\right)}$
and,~for MSSM,
\par
\noindent
\ba 
F_{\tau}&=&{-h_\tau^2}/{\left(16\pi^2\cos^2\beta\right)},\nonumber\\
F_u&=&\left(1\over{16\pi^2}\right)\left({6\over5}g_1^2+6g_2^2-
6{h_t^2\over\sin^2\beta}\right),\label{eq6}
\ea
\par
\noindent
but, for SM,
\par
\noindent
\ba 
F_{\tau}&=&{3h_\tau^2}/\left( 32\pi^2\right),\nonumber \\
F_u&=&\left(3g_2^2-2\lambda-6h_t^2-2h_\tau^2\right)/\left(16\pi^2\right).
\label{eq7}
\ea
\par
\noindent
Natural occurence of a SUSY scale near $300-1000$ GeV does not permit
radiative magnification below this scale where the mixinq angles
remain constant in the presence of the SM.\\
The resulting RG evolutions of quark and neutrino mixings  in different cases
have been shown in \cite{mpr-PASCOS:2003,mpr-PRD:2003,mpr-others,ampr:2007,mpr-split:2005}

 For such high scale mixing unification (HUM), the desired   QD neutrino mass scale is $m_i\simeq m_0 >  0.15 (i=1,2,3)$ eV. WMAP data \cite{wmap-1} suggest the bound $\Sigma_C \equiv \sum m_{\nu_i} < 0.69 $ eV but, consistent with priors, it has been also noted that $\Sigma_C \le 1$ eV \cite{Hannestad}.  
Although  recent Planck satellite data has determined a cosmological bound
$\Sigma_C \leq 0.23 $ eV \cite{Planck15} and ( or even  lower \cite{Sunny:2016}), the same data  have been noted to admit $\Sigma_C \le 0.71$ eV in the absence of $\Lambda$CDM based theory of the Universe \cite{Planck15}. However 
all neutrino mass values needed for HUM \cite{mpr-PRD:2003,mpr-others,ampr:2007,mpr-split:2005} are in concordance  with the most recent laboratory bound from KATRIN collaboration \cite{KATRIN} that has reached the limit $ m_{\nu} < 1 $ eV. 
QD neutrino masses needed elsewhere
 \cite{BLP:1993,Chankowski:1993,Balaji-PRL:2000,Balaji-PLB:2000,Balaji-PRD:2001,HGM:2001,Caldwell:1994} are also allowed by KATRIN results.  It has been further noted that the presence of SUSY substantially below the GUT-Planck scale is not
a necessary criteria for understanding such RG origin of neutrino
physics as the mechanism works profoundly even with very high scale split-SUSY \cite{mpr-split:2005}.
\section{Advantages of SO(10)}\label{sec:so10}
The discussions stated below apply to both SUSY or non-SUSY SO(10) or $E_6$.
The SU(5) GUT predicts  ${\bf 15}$ SM fermions of one generation in two different
representations ${\bar 5}_F+ {10}_F$, but they are all unified with
right-handed (RH) neutrino (N) into a single spinorial representation
${16}_F$ in SO(10).  Dirac neutrino mass generation in SM or SU(5)
needs introduction of N externally, but in SO(10) this follows automatically
 from Yukawa
interaction $Y{16}_F{16}_F{10}_H$ where 
${10}_H \supset \phi (2,1/2,1)$ which is the standard Higgs scalar
doublet. 
\subsection{D-Parity Breaking and Emergence of New SO(10) Models}\label{sec:dpar}
Before 1984 the breaking of left-right discrete symmetry ($\equiv$
Parity(P) = space-inversion symmetry) was synonymous with $SU(2)_R$
breaking \cite{ps:1974,RNM-JCP:1975}. This did not permit low mass $W_R$ bosons or $SU(4)_C$ \cite{ps:1974} breaking scales
accessible to accelerators, although a two-step breaking of left-right symmetric gauge theory was shown to predict a low-mass $Z^{\prime}$ boson \cite{mkp-cch:1983} in concordance with $K_L-K_S$ mass difference. The discovery and identification of
D-Parity properties of SO(10) \cite{cmp-PRL:1984,cmp-PRD:1984} representations paved the ways for
lowering such mass scales substantially leading to new classes of
SO(10) accessible to experimental tests \cite{cmgmp:1985}.
 It was at first noted \cite{cmp-PRL:1984} that if the LRS theory $G_{2213D}$
 has a scalar singlet $\sigma$ that is odd under the L(left) $\to$ R (right) transformation
, then its VEV $\langle \sigma \rangle{=V_\sigma}$ would
break the LR discrete symmetry without breaking the gauge symmetry
leading to
$G_{2213D}\to G_{2213}$ (or $G_{224D}\to G_{224}$). More important is the identification of such
scalar singlets in $SO(10)$.  
Defining D-Parity  as an element of SO(10) gauge tranformation that
takes a fermion $\psi_L\subset {16}_F$ to its conjugate
$\psi_L^C(\propto \psi_R^*)$
which is also in
the same ${16}_F$, invariance under $D$ guarantees 
left-right (LR) discrete symmetry with $g_{2L}=g_{2R}$, but spontaneous breaking of $D$ implies
breaking of LRS with $g_{2L}\neq g_{2R}$ but without breaking the
gauge symmetry $G_{224}$ or $G_{2213}$. 
The D-Parity properties of SO(10) scalar components were identified for the first time
\cite{cmp-PRL:1984,cmp-PRD:1984}. 
Considering branching rules \cite{Slansky:1981} of SO(10) scalar representations under Pati-Salam
symmetery ($G_{224D}$) 
\par\noindent{\bf {SO(10) $\supset SU(2)_L\times SU(2)_R\times
    SU(4)_C(g_{2L}=g_{2R})(\equiv G_{224D})$}:}
\ba
{10}&=&(2,2,1)+(1,1,6), \nonumber\\
{16} &=&(2,1,4)+(1,2,{\bar 4}), \nonumber\\
{45}&=&S_o(1,1,15)+(3,1,1)+(1,3,1)+(2,2,6), \nonumber\\
{54}&=&S_e(1,1,1)+(3,3,1)+(2,2,6)+(1,1,20),  \nonumber\\
{126}&=&(2,2,15)+\Delta_L(3,1,10)+\Delta_R(1,3,{\bar {10}})+(1,1,6),\nonumber\\
{210}&=&S_o(1,1,1)+S_e(1,1,15)+(2,2,10)+(2,2,{\bar {10}}) \nonumber\\
&+&(3,1,15)+(1,3,15)+(2,2,6),\label{eq:dh10}
\ea
The $G_{224D}-$ singlet
  $(1,1,1)$ in ${54}_H({210}_H)$  were identified to be
  D-even (D-odd) leading to SO(10)$\to G_{224D}(G_{224})$ through
  its GUT scale VEV. Similarly the neutral component of $ (1,1,15)_H$ in ${45}_H({210}_H)$ was identified to
  be D-odd(D-even) leading to  $G_{2213}(G_{2213D})$. In
  this manner a large number of new symmetry breaking chains were
  predicted \cite{cmp-PRL:1984,cmp-PRD:1984,cmgmp:1985} with interesting
  phenomenological consequences \cite{cmp-PRL:1984,cmp-PRD:1984,cmgmp:1985,RNM-mkp:1993,lmpr:1995,Pal:1994,mkp-cch:1989,mkp-pkp-PL:1991} including gauge coupling unification. 
Before the D-Parity properties of ${210},{45}_H,{54}_H$ were known, LR discrte
symmetry breaking at the GUT scale was employed through fine
tuning  assumption predict left-right asymmetric gauge theory $G_{224} (g_{2L}\neq g_{2R})$ at $M_C\simeq 10^6$ GeV leding to bservable $n-{\bar n}$ oscillation and TeV scale $Z_R$ boson \cite{mkp:1983}.
\section{Theorems  on Vanishing Uncertainties in  GUTs}\label{sec:theo}
 In a GUT, besides the RGE corrections due to running
  gauge couplings, there are other uncertainties due to GUT
 threshold corrections \cite{Weinberg:1980,Hall:1981,mkp:1987,Dixit:1989}, Plank scale induced higher dimensional
 operators \cite{mkp-pkp:1991,mkp-pkp:1992,mkp:1998}, and string threshold effects \cite{strth} on the model
 predictions of
$\sin^2\theta_W$ or $G_{224D}$ breaking intermediate scale
 {$M_I=M_P$}. This was also projected as a major source of uncertainty
 in $\sin^2\theta_W$ prediction with Pati-Salam intermediate
 breaking \cite{Dixit:1989}. In sharp contrast, the 
 following three
theorems were discovered to predict  complete absence of such  uncertainties establshing profoundly predictive nature of SUSY and non-SUSY GUTs with $G_{224D}$ intermediate symmetry. 
 
\par\noindent{\bf (1). Theorem-1:}\cite{mkp-pkp:1991} Whenever a grand unified theory
possesses the gauge symmetry  
  $SU(2)_L\times SU(2)_R\times SU(4)_C\times D (=G_{224D},
  g_{2l}=g_{2R}))$  at the highest intermediate scale ($M_I$), the
  one-loop GUT threshold contribution to $\sin^2\theta_W(\mu), (\mu \ge M_I) $ by every
  class of superheavy particles (gauge bosons, Higgs scalars and
  additional fermions) vanishes. The result also applies with
  supersymmetry, infinite towers, or higher dimensional operators, and
  is independent of other intermediate symmetries at lower scales \cite{mkp-pkp:1991}. 
\par\noindent{\bf (2). Theorem-2:}\cite{mkp-pkp:1992} In all symmetry breaking chains where
  the symmetry $G_{224D}$ occurs at the highest intermediate scale
  $M_I$, all higher order multi-loop corrections on  $\sin^2\theta_W(\mu)$
are absent in the
  mass range $\mu=M_I- M_U$. This theorem also holds with
  supersymmetry or string inspired models \cite{strth}.  
 \par\noindent{\bf (3). Theorem-3:}\cite{mkp:1998} In all symmetry breaking chains where
  $G_{224D}$ occurs at the highest intermediate scale $M_I$, the 
     scale $M_I$ has vanishing contributions  from all sources of
     corrections arising at mass scales $\mu > M_I$ \cite{mkp:1998}.\\
  These corrections also includes those due to gravitational or Planck
  scale effects due to higher dimensional operators and/or string
  threshold effects.
Consequently, $M_I$ prediction of a SUSY SO(10) has been shown to be
unaffected by the number of ${16}_H\oplus {\ovl {16}}_H$ pairs above
$\mu=M_I$ although it has the capability to change the value of
$M_I$  only through lighter components in ${16}_H\oplus {\ovl {16}}_H$
and/or ${45}_H$. One example of solutions is $M_I=10^{12.5}$ GeV and
$M_U=10^{17.6}$ GeV indicating a very stable proton and string scale
unification \cite{mkp:1998} with GUT fine-structure constant well below
the perturbative limit. All the fields used in this SUSY SO(10) belong
to the string compacification model \cite{strth}.
 
As a result of these  theorems, it is interesting to examine SUSY and non-SUSY
SO(10) model predictions:
\be
E_6 \,\,{\rm or}\,\, SO(10) \to G_{224D} \to  SM \label{eq:pschain}
\ee
In non-SUSY case, the intermediate scale remains fixed at
$M_I \simeq 10^{13.6}$ and $M_{GUT}=10^{14.8}$ GeV. At first sight
the GUT scale appears to be
volnurable to Super Kamiokande  limit on proton lifetime
\cite{SuperK} for $p\to e^+\pi^0$. But the theorems also come to rescue. By fine-tuning when the
scalar multiplet $\xi (2,2,15) \subset {126}_H$ is placed at $M_I$, the theorem predicts no change in the values of $\sin^2\theta_W$ or $M_I$ from the minimal case. But   the RG effects do 
increases the unification scale leading to  $M_{GUT} > 10^{15.5}$ GeV which easily evades the Super Kamiokande limit.
In SUSY case $M_I\simeq 10^{14}$ GeV and $M_U\simeq 10^{16.5}$ GeV \cite{mkp:1998}.\\
A theorem
has been also proposed which is valid for threshold corrections due to Higgs representation
that does not acquire VEV and has all degenerate  components.
\par\noindent{ \bf(4). Theorem-4:} \cite{RNM:1992} Threshold corrections to unification scales, $\sin^2\theta_W$  and ${\alpha}_S$ vanish
for a Higgs multiplet of grand unification group which does not acquire VEV if we make
the plausible assumption that all its submultiplets are degenerate in mass after symmetry
breakings.\\
\par\noindent{\bf (5). Vanishing Planck scale effect on SUSY $M_U$ :} \cite{ppdc:2003}
Planck scale effects due to $5-dim.$ operators are known to affect the GUT scale predictions substantially in a grand unified theory. But
it has been shown that in SUSY SO(10) breaking to $G_{2213}$ the D-Parity even and odd combinations arising from ${210}_H$ cancel out the effects of the two non-renormalisable corrections on $M_U$.
 
\section{Leptogenesis Within Gravitino Constraint}\label{sec:gr10}
 In the RHN extended SM the reheating temperataure after inflation is to be as
high as $T_{rh}\simeq 10^8-10^9$ GeV \cite{Davidson} leading to overproduction of gravitinos
  \cite{Khlopov:1986} and depletion of 
  deuterium relic abundance below acceptable limits. Another
  draw back of high type-I seesaw scales is that the proposed mechanism
can neither be directly verified in near future, nor can it be
disproved. They also predict negligible LFV
decay branching ratio (Br.)  for $l_{\alpha}\to l_{\beta}\gamma (\alpha\neq
\beta=e,\mu,\tau)$ and $\mu\to e{\bar e}e$ which have expermental
limits $Br.\simeq 10^{-9}\to 10^{-13}$.

These issues have been addressed in the the  SUSY SO(10) breaking models 
 \cite{majee-mkp-arc:2007,mkp-arc:2010}  
\begin{eqnarray}
SO(10) & \stackrel {(M_U)}{\longrightarrow} & [{\cal
 G}_{2213D}] 
\stackrel {(M_P)}
{\longrightarrow}  [{\cal G}_{2213}] \nonumber \\
&
\stackrel {(M_R)}
{\longrightarrow}&[{\cal G}_{213}] \stackrel {(M_Z)}{\longrightarrow} SU(3)_C \times U(1)_Q \;\;.\label{lep-chain}
\end{eqnarray}
\par
The first stage of spontaneous symmetry breaking (SSB) is carried
out by assigning GUT scale vacuum expectation values to the
$\Phi_{54}$ of $SO(10)$ along the direction of a
singlet under $G_{224}$ and $G_{2213D}$. The second step os SSB occurs
when $(1, 1, 15)$ under $G_{224}$ contained in ${210}_H$ gets
$VEV\simeq M_P$. At
this stage D-parity remains intact with equal LR gauge couplings of
$SU(2)_L$ and $SU(2)_R$, $g_{2L}=g_{2R}$ \cite{cmp-PRL:1984}. The
second stage of SSB takes place by assigning vacuum expectation
value to the D-Parity odd singlet also contained in
$\Phi^{(2)}_{210}$ of $SO(10)$.  By suitable fine tunings of
the trilinear couplings beteen ${\bf 210}$ and the ${\bf {126}_H
\oplus \overline {126}_H}$ or ${\bf {16}_H \oplus \overline
{16}_H}$ the right handed triplets ${\bf \Delta_R \oplus
\overline {\Delta}_R} \subset {\bf {126}_H \oplus \overline
{126}_H}$ and the RH doublets ${\bf \chi_R \oplus \ovl {\chi}_R}
\subset {\bf {16}_H \oplus \ovl {16}_H}$ are made much lighter
compared to their left-handed counterparts. By adopting higher
degree of fine tuning for the RH triplet compared to the RH
doublet, the components of the RH triplet pairs can be assigned
masses between 100 GeV to a few TeV while the RH doublet pairs
are kept heavier, but sufficiently lighter than the GUT scale.
Although we do not assign any  VEV directly to the neutral
components of the RH-triplets in ${\bf {126}_H \oplus \overline
{126}_H}$, we find that the assigned VEV
 of the RH-doublet in ${\bf {16}_H}$, automatically induces the RH-triplet VEV.  Smaller is the
RH-triplet mass fixed by the D-parity breaking mechanism, larger
is the induced triplet VEV.

This symmetry breaking gives the  Yukawa  Lagrangian near the
intermediate scale      
\ba
{\cal L}_Y = Y \ovl \psi_L \psi_R \Phi + f\psi^T_R\tau_2\psi_R
\bar {\Delta}_R + F\ovl \psi_RS\chi_R + \mu S^TS+H.c.\label{yuklm}
\ea
 where $\psi_{L, R}$ are left- (right-) handed lepton doublets. In
the $(\nu, N, T)$ basis this  leads to a $3\time 3$ mass matrix
\begin{equation}
M_\nu = \left( \begin{array}{ccc}
0 & m_D & 0  \cr
m_D^T & M_N & M_X \cr 
0 & M_X^T & \mu  \cr
\end{array}\right ).  \label{matrix}
\end{equation}
Here the $N-S$ mixing matrix arises through the {\em vev} of the
RH-doublet field with $M_X = Fv_{\chi}$, where
$v_{\chi}=\langle \chi^0_R\rangle $, and the RH-Majorana neutrino
mass is generated by the induced {\em vev} of the RH-triplet
with $M_N=fv_R$,  with $v_R=\langle
\overline{\Delta}^0_R\rangle$. The {\em vev} of the weak
bi-doublet $\Phi(2,2,0,1) \subset 10_H$ of $SO(10)$ yields
the Dirac mass matrix for neutrinos, $m_D = Y \langle\Phi^0
\rangle$.

The  block diagonalization of this mass
matrix results in a cancellation among  the  Type-I see-saw
contributions and  the light neutrino mass $m_{\nu}$  is
dominated by the inverse see-saw,
\ba
m_{\nu} &=&~ -m_D~[M_X^{-1}\mu (M_X^T)^{-1}]~m_D^T ,
\label{inv1}
\ea
\vskip -30pt
\ba
   M_T &=& ~\mu -M_X~M_N^{-1}~M_X^T , 
\label{inv2}
\ea
\vskip -30pt
\ba
M &=& ~M_N + ~{M_X}~M_N^{-1}~M_X^T.
\label{inv3}
\ea
In this model the type-II seesaw contribution is negligible \cite{mkp-arc:2010}.

\par
Assuming diagonal basis for RHN, $M_N= {\rm
{diag}} (M_{N_1}, M_{N_2}, M_{N_3})$, the model  generates $N_i-S_j$ mixing
angles,
\be  
\sin \xi_{ij} \simeq {M_{X_{ij}} \over M_{N_i}}. 
\label{mixxi} 
\ee
  
\begin{figure}[h!]
\begin{center}
\includegraphics[scale=0.3]{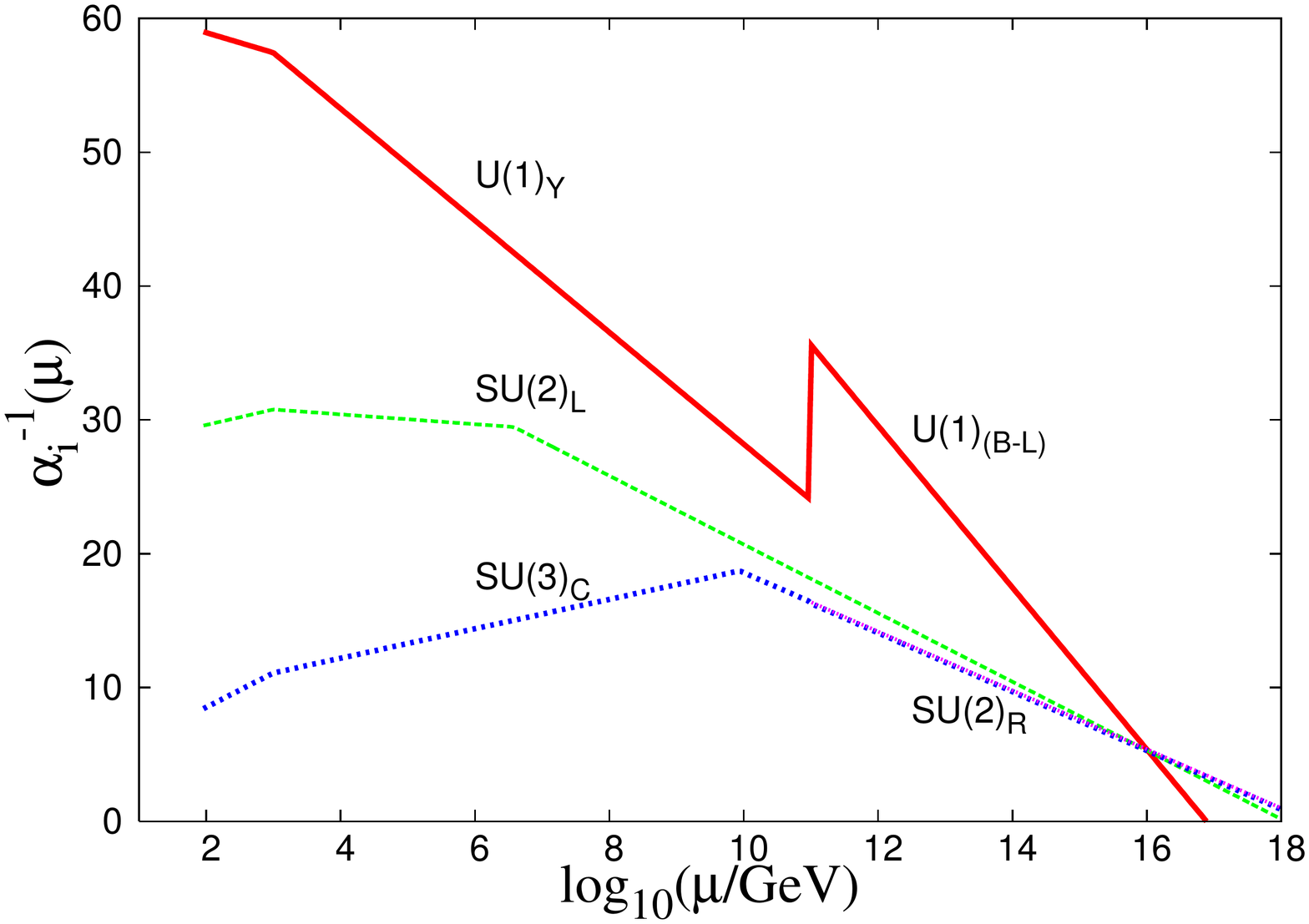}
\end{center}
\caption{ Evolution of gauge couplings leading to unification at the
  SUSY SO(10) GUT scale
  $M_{U}=10^{16}$ GeV.}
  \label{fig:gi10}
\end{figure}

Purely from SUSY $SO(10)$ considerations, the method of keeping  the relevant Higgs
scalars substantially lighter than the GUT scale needed for pecision
coupling unification has been
discussed in  \cite {mkp:2008}. As a typical example, the evolution of the  gauge
couplings and unification at the GUT scale are shown in
Fig. \ref{fig:gi10} that led to the solution
\ba 
M_R = 10^{11} ~{\rm GeV}, ~~ M_U = 10^{16} ~{\rm GeV}, \label{soln}
\ea 
with $\alpha_G^{-1}=5.3$ which is  well within the perturbative
limit.  In  Fig. \ref{fig:gi10} the couplings for $SU(2)_R$ and
$SU(3)_C$ are found to be almost ovelapping above the scale $M_R$
because of a fortuitous identity of their respective beta function
coefficients and near equality of the boundary values at $M_R$ in
this example. The change in slopes at $M_\sigma$ and $M_C$ are
clearly noticeable.

\begin{figure}[h!]
\begin{center}
\includegraphics[scale=.5]{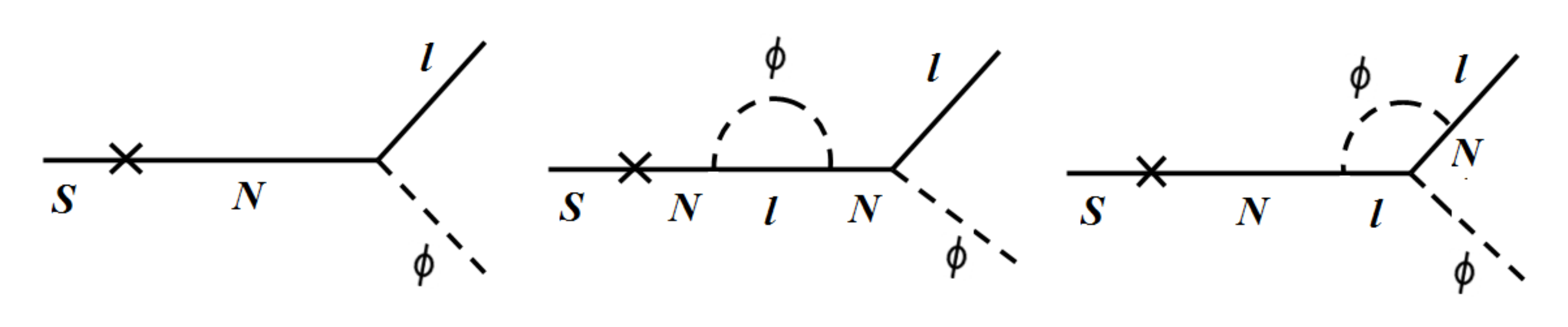}
\end{center}
\caption{ The siglet fermion decay leptogenesis in SUSY SO(10)}
  \label{fig:sd}
\end{figure}

Proton lifetime in this model is cosistent with Super Kamiokande limit \cite{SuperK}   
  As pointed out neutrino masses and mixings are fitted by inverse
  seesaw mediated by a singlet which also generates lepton asymmetry
  through its decay as shown in
Fig. \ref{fig:sd}.

 The formula for the singlet fermion decay rate assumes the form ,
\ba
\Gamma_{S_1} = \frac{1}{8\pi }M_{S_1}  \frac{K_1}{K_2} & \left[(|{\tilde
U}_{11}|)^2 \sin^2\xi_{11}
(Y_D^{\dagger}Y_D)_{11}
+(|{\tilde U}_{12}|)^2\sin^2\xi_{32}(Y_D^{\dagger}Y_D)_{33}
\right. \nonumber\\
&\left. +(|{\tilde
U}_{13}|)^2\sin^2\xi_{23}(Y_D^{\dagger}Y_D)_{22} \right],
\label{gammas1}          
\ea
where $K_1$,  $K_2$ are modified Bessel functions. 
Even though $Y_D$ is of the same order as the up-quark
Yukawa matrix, the smallness of $\Gamma_{S_1}$, compared to the Type-I
see-saw case, originates from two sources:
(i) Allowed values of $M_{S_1}\ll M_{N_i} (i=1,2,3)$, (ii)
$\sin^2{\xi}_{jk} \ll 1 ~(j, k = 1, 2, 3)$. These two features achieve
the out-of-equilibrium condition at temperature $\sim M_{S_1}$
satisfying the gravitino constraint.  
A compact formula for CP-asymmetry has been also derived as a function
of model parameters \cite{mkp-arc:2010}

Defining $n_{B}$ as the net baryon number density over anti baryons
and $n_{\gamma}$ as photon number density, the
estimated baryon asymmetry turns out to be
\begin{equation}
\eta_B \equiv  {n_B\over n_{\gamma}}   \simeq 10^{-2}\kappa\epsilon_1.
\label{bauexp}
\end{equation}
in good agreement with  \cite{Planck15}:
\be
(\eta_B)_{\rm expt} = (6.15 \pm 0.25)\times 10^{-10}.
\label{baudat}  
\ee
\section{Flavour Unification and Fermion Masses  Through SUSY SO(10)$\times
  S_4$}\label{sec:lrsrp}
A number of attempts exist to explain neutrino masses as well as
chargrd frmion masses using flavour symmetries
\cite{Adhikary:2006}. In
this section we discuss briefly how conservation of both the
symmetries, D-Parity and R-Parity, guarantees an intermediate scale and all the fermion mass
fittings  through SUSY SO(10)$\times S_4$ flavour symmetry \cite{mkp:2008}. 
 SUSY SO(10) predicts R-Parity ($R_p={(-1)}^{3(B-L)+2S}$) as its intrinsic gauged discrete symmetry for the stability of dark matter (wino, bino,neutralino etc.) whenever spontaneously broken 
through ${126}_H\oplus {\overline 126}_H$. It has also the ability to predict space-time
left-right discrete symmetry as a remnant of continuous gauge symmetry
(=D-Parity) to survive down to lower intermediate scale.  However  minimal SUSY SO(10)
models with interesting predictiok of type-I $+$ type-II seesaw is
known to forbid intermediate gauge symmetry breaking  although, as we
have seen in the previous section, lighter degrees  of freedom
resulting from fine tuning do permit intermediate left-right gauge symmetry with
  $M_R << M_{GUT}$. 
We consider $S_4$ flavor symmetry for three fermion generations  through  $SO(10)
\times S_4$ \cite{hagedorn} in  the following symmetry breaking model
{\Large $SO(10)\times S_4 \mathop{{\rightarrow}^{210}_{M_U}} ~~G_{2213}\times S_4$}  
{\Large $\mathop{{\rightarrow}^{126+\ol {126}}_{M_R}}~~G_{213} 
\mathop{{\rightarrow}^{10}_{M_W}}~~U(1)_{em}\times SU(3)_C$}.\\
The representation content of the $SO(10)
\times S_4$ theory is shown in Table \ref{tab1}.
\begin{table*}
\caption{Particle content of the model and their 
transformation properties under $S_4\times SO(10)$   }   
\begin{tabular}{lcccccccc}\hline
$\bf {Fermions}$&& $\bf {Higgs ~Bosons}$&&&\\\hline
${\bf \Psi_i, (i=1,2,3)}$&${\bf S}$&${\bf \Phi }$&
${\bf A_{1,2,3}}$&$\bf {\overline {\Sigma}_0 \oplus {\Sigma}_0 }$&
$\bf  {H_0}$&$\bf {H_{1,2}}$&$\bf {H_{3,4,5}}$\\
\hline\\
${\bf {3^{\prime}}}\times {\bf {16}}$&${\bf {1\times 54}}$&${\bf {1}}\times {\bf {210}}$&${\bf { 3\times 45}}
$ &
${\bf {1}}\times {\bf \overline {126}\oplus 126}$&${\bf {1}}\times {\bf {10}}$&${\bf {2}}
\times {\bf {10}}$&${\bf {3}}\times {\bf {10}}$\\\hline
\end{tabular}
\label{tab1}
\end{table*}  
In this model the $G_{2213}$ representations which have masses at the
intermediate scale are
\ba
\Dl(3,1, -2, 1)\op \Dr(1,3 , -2, 1)\op 
\ol {\Dl}(3,1, 2,1) \op  \ol {\Dr}(3,1, 2, 1),\nonumber \\
6(2,2,0,1), 3(1,1,0,8),\label{eqps} 
\ea
where ${\bf 6=3+2+1}$, and ${\bf 3,2}$ and ${\bf 1}$ are triplet, doublet, and  singlet ,
respectively, under $S_4$. These  result in the
respective beta-function 
coefficients in the mass range \\
\par
\begin{figure}[h!]
\begin{center}
\includegraphics[scale=0.3]{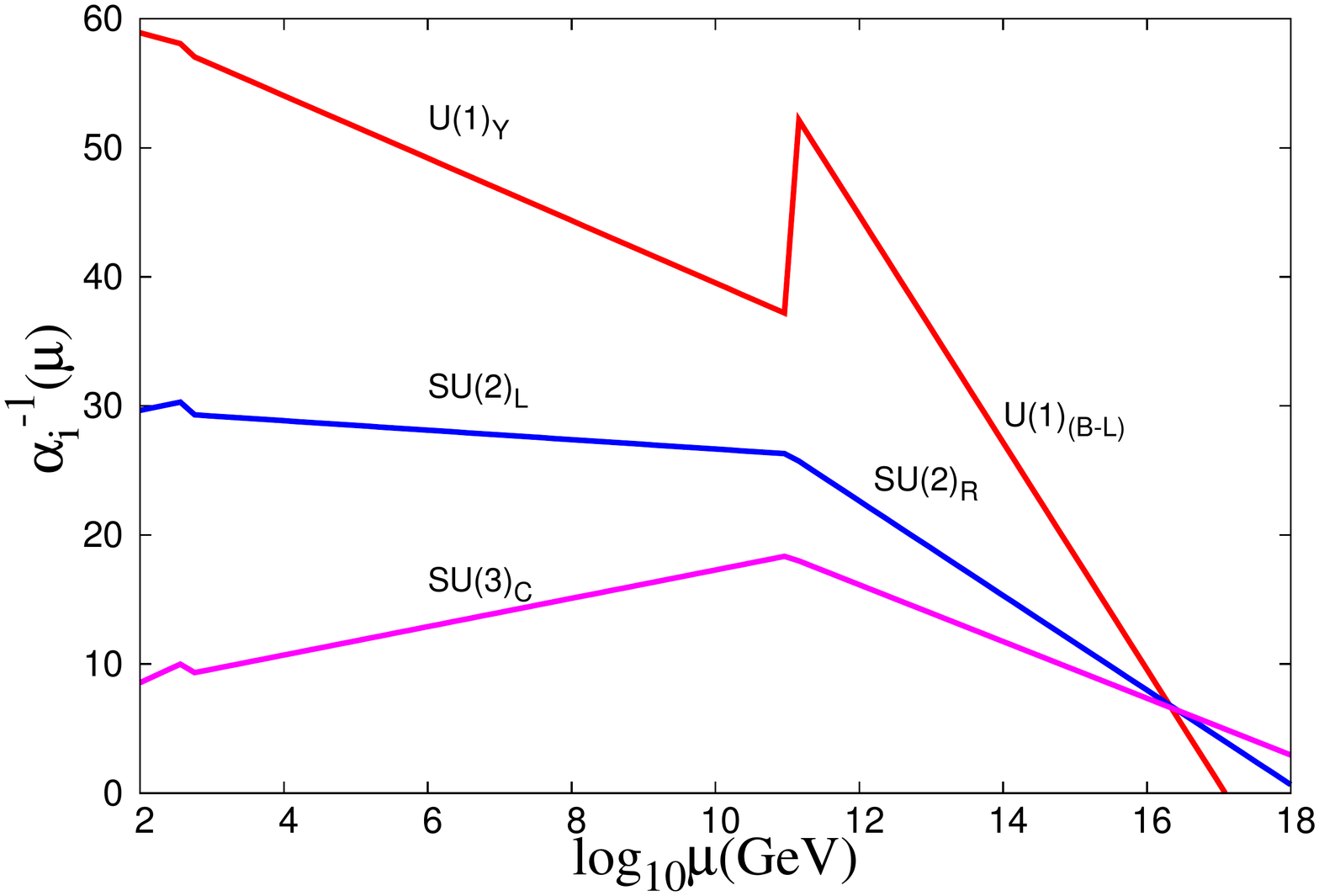}
\end{center}
\caption{ Evolution of gauge couplings leading to unification at the
  SUSY $SO(10)\times S_4$ GUT scale  $M_{U}=2\times 10^{16}$ GeV.}
  \label{fig:S42}
\end{figure}

\noindent {\bf {$\mu =  M_R - M_U$:}}\\
\be
a'_{BL} = 24, ~a'_{2L} =a'_{2R} =10,  ~a'_{3C} = 6 .\label{eqfr}
\ee 
 We find that 
values of the left-right symmetry breaking  scale $M_R$ are permitted 
over a wide range,
\be
5\times 10^{9} ~{\rm GeV} \le  ~M_{\rm R}~  \le ~10^{16} ~{\rm GeV}.  
\label{bound}
\ee 
but having almost the same value of unification scale $M_U = 2\times 10^{16}$ GeV 
for all solutions. One example with $10^{13}$ GeV is
  shown in Fig. \ref{fig:S42}.
This model has capability to fit all fermion masses and mixings
including  neutrino data \cite{mkp:2008} by Type-I seesaw. The predicted proton
lifetime is about $1-2$ orders longer than the Super Kamiokande limit \cite{SuperK}.   
\section{Other Applications in SUSY SO(10)}\label{sec:moreSUSY}
SUSY SO(10) model building with ${45}_H$, or ${54}_H$, or ${210}_H$ combined with ${126}_H\oplus {\overline {126}}_H$ has a number of attractive predictions in neutrino physics, cosmology and all charged fermion mass fitting. It predicts type-I$\oplus$ type-II ansatz that fits oscillation data. The heavy RHN's mediating type-I or LH scalar triplet mediating type-II are capable of explaining baryon asymmetry of the universe via leptogenesis. Such SO(10) breaking predicts R-Parity as gauged discrete symmetry that guarantees stability of dark matter. Despite these attractions, the SUSY SO(10) theory starts  becoming non-perturbative \cite{Aulakh:1998} at mass scales few times larger than the GUT scale $\mu \ge (\rm few)\times 2 \times 10^{16}$ GeV. A resolution of this difficulty has been suggested via GUT threshold effects \cite{Cajee-mkp:2005} ensuring perturbativity till the Planck mass. \\ 
Currently Starobinsky \cite{Starobinsky} type inflation appears to
describe the big-bang comology most effectively. Using the identified D-Parity
properties of SO(10) such an inflationary picture has been realised in
SUSY SO(10)\cite{Ellis:2016} with double seesaw ansatz for neutrino
masses and verifiable proton lifetime predictions in near future. With
D-Parity
broken at the GUT scale,
SUSY SO(10) predictions of low-mass $W_R$ bosons, proton decay,
inverse seesw, and leptogenesis have been also investigated \cite{Valle:2005}.
\section{Non-SUSY GUTs Confronting Particle Physics Issues
}\label{sec:nsgut}
\subsection{Intermediate Breaking Models}
\begin{figure}[h!]
\begin{center}
\includegraphics[scale=0.5]{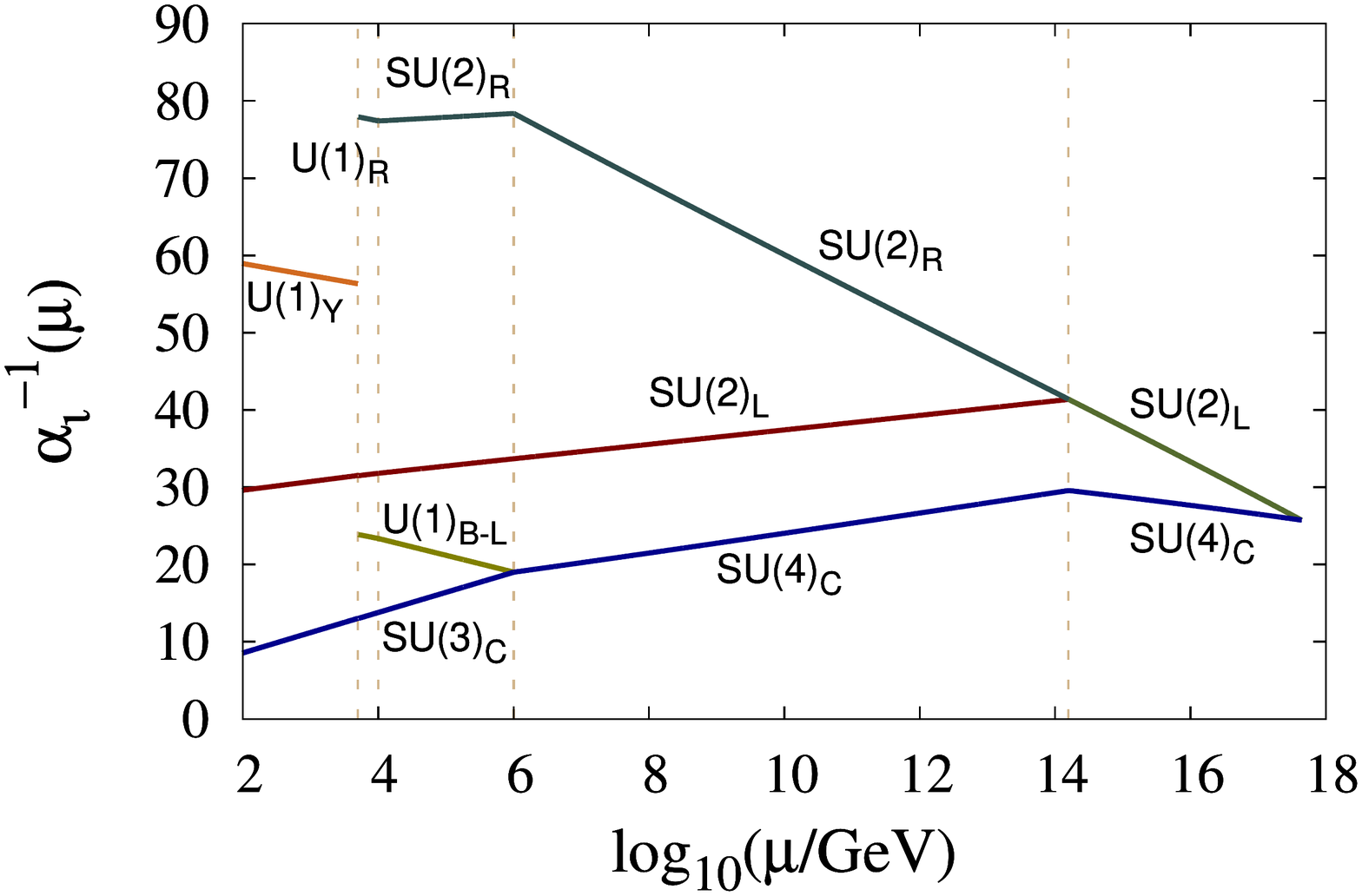}
\end{center}
\caption{ High scale unification in Non-SUSY SO(10) with prospects for
  verifiable $W_R,Z_R$ bosons, LFV and LNV decay.}
  \label{fig:WRZR}
\end{figure}

Even before emergence of MSSM unification, especially from CERN-LEP data \cite{Amaldi:1991}, it was noted that SUSY may not be a compelling requirement for unification. This was worked out in detail in  non-SUSY SO(10) with one, two, or mopre intermediate gauge symmetries \cite{cmp-PRL:1984,cmp-PRD:1984,cmgmp:1985}. Two minimal examples with $G_{2213}$  or $G_{2213D}$ and  $G_{224}$ or $G_{224D}$ intermediate gauge symmtries and others have been discussed including threshold effects \cite{RNM-mkp:1993,lmpr:1995,mkp-pkp-PL:1991}.\\ 
Generally high scale seesaw  models are not directly verifiable; they
also predict negligible LFV branching ratios. However, it has been
found \cite{app:2013,pas:2014} that if  $G_{224D}$ occurs at higher scale, $G_{224}$ symmetry
can survive down to $10^5$ GeV. In such a model the $G_{2213}$
breaking may occur at TeV scales leading to $W_R,Z_R$ bosons
accessible to LHC. The model also has capability to predict  LFV decay
branching ratios only about $2-3$ orders lower than the current
experimental limits. Neutrinoless double beta decay is predicted to be
accessible by ongoing experimental searches even for normally ordered
(NO) or invertedly ordered (IO) neutrino mass hierarchies as the LNV decay
process is predicted by a low mass sterile neutrino of mass $\sim 10$
GeV which is found to be a generic feature with Higgs representations ${126}_H\oplus {16}_H$ as noted below.
Gauge coupling unification in this model \cite{app:2013} is shown in Fig.\ref{fig:WRZR}.
\subsection{Verifiable $W_R,Z_R$, Inverse
  Seesaw, LFV, and $(\beta\beta)_{0\nu}$ with $G_{2213}$ Intermediate Breaking }
Interestingly, if Planck scale effects are utilised \cite{mkp-pkp-PL:1991}, non-SUSY SO(10)
with D-Parity broken at the GUT scale having only
 the lone $G_{2213}$ intermediate symmetry gurantees verifiable
 $W_R,Z_R$ bosons with masses $1-10$ TeV, neutrino masses by inverse
 seesaw, experimentally accessible lepton flavour violating branching
 ratios, and neutrinoless double beta decay close to the current
 experimental limits even with hirarchical neutrino masses in
 concordance with cosmological bounds \cite{ap:2011,bpn-mkp:2013,bs-mkp:2015,bs-mkp:2017}. Other recent applications with D-Parity broken intermediate gauge symmetry have been dicussed in \cite{int-DM}. 
\subsection{Unification Without Intermediate Gauge Symmetry}
\subsection{Hybrid Seesaw, Dark Matter and Leptogenesis} 
\begin{figure}[h!]
\includegraphics[scale=0.3]{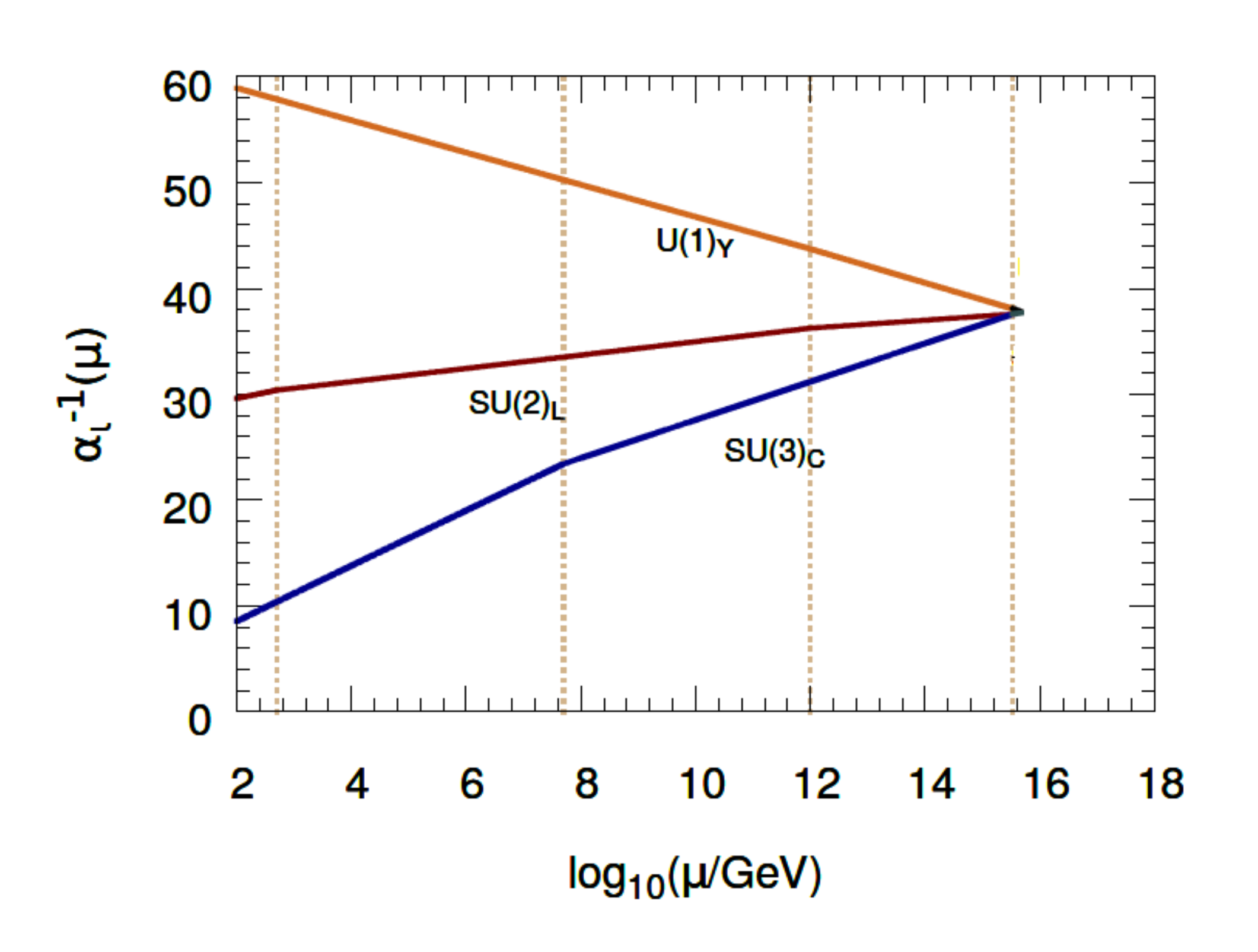}
\includegraphics[scale=0.7]{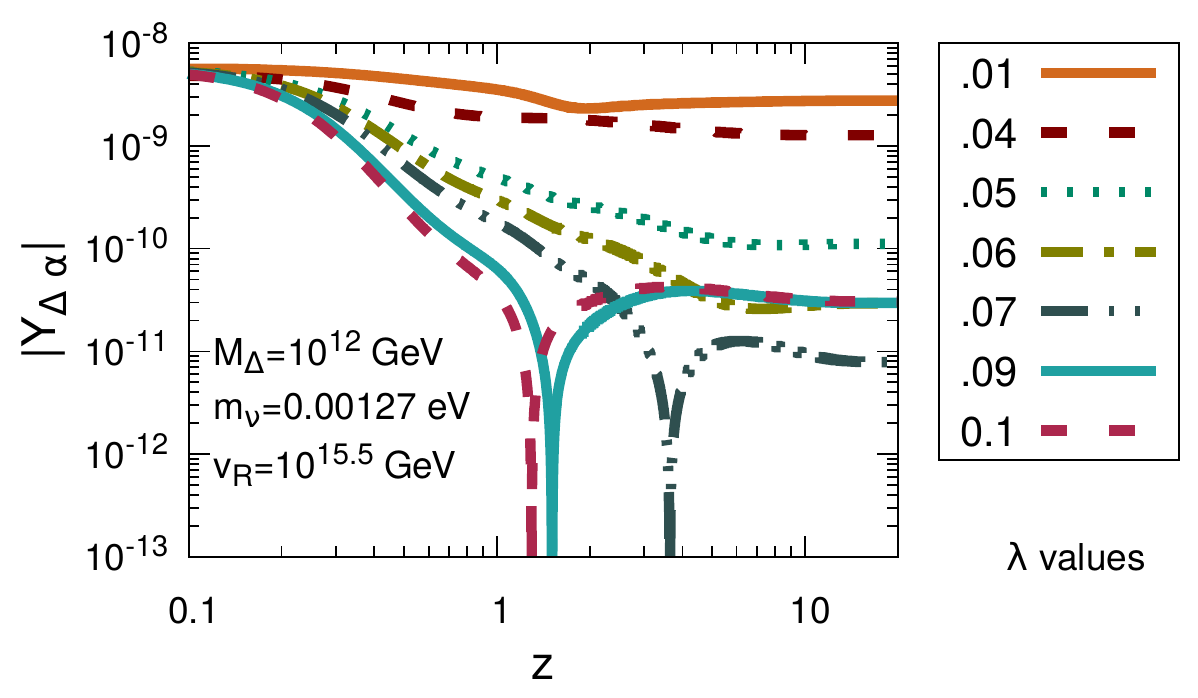}
\caption{Evolution of gauge couplings in the hybrid seesaw model of SO(10) \cite{pnsa:2017} as shown
  in the left panel. Prediction of baryon asymmetry as shown in the right 
panel.}
 \label{fig:hybr}
\end{figure}

Without using any intermediate gauge symmetry but using only few lighter fields precision gauge coupling in one example \cite{pnsa:2017} is shown in the left panel of Fig.\ref{fig:hybr}. Baryon asymmetry prediction of this model has been shown in the right panel of the
same Fig. \ref{fig:hybr}.

\subsection{Minimally Modified Grand Deserts}
\begin{figure}[h!]
\includegraphics[scale=0.4]{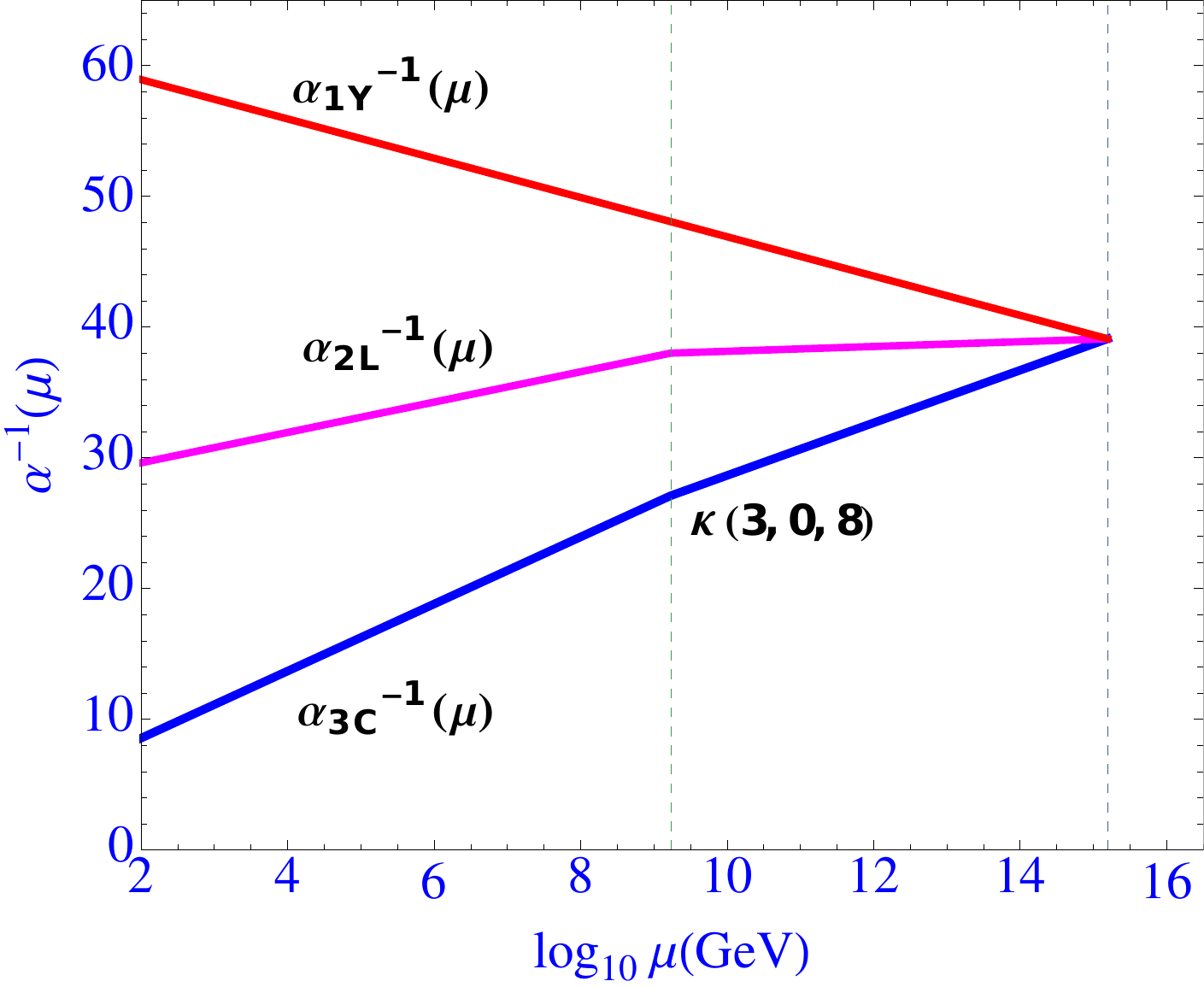}
\includegraphics[scale=0.4]{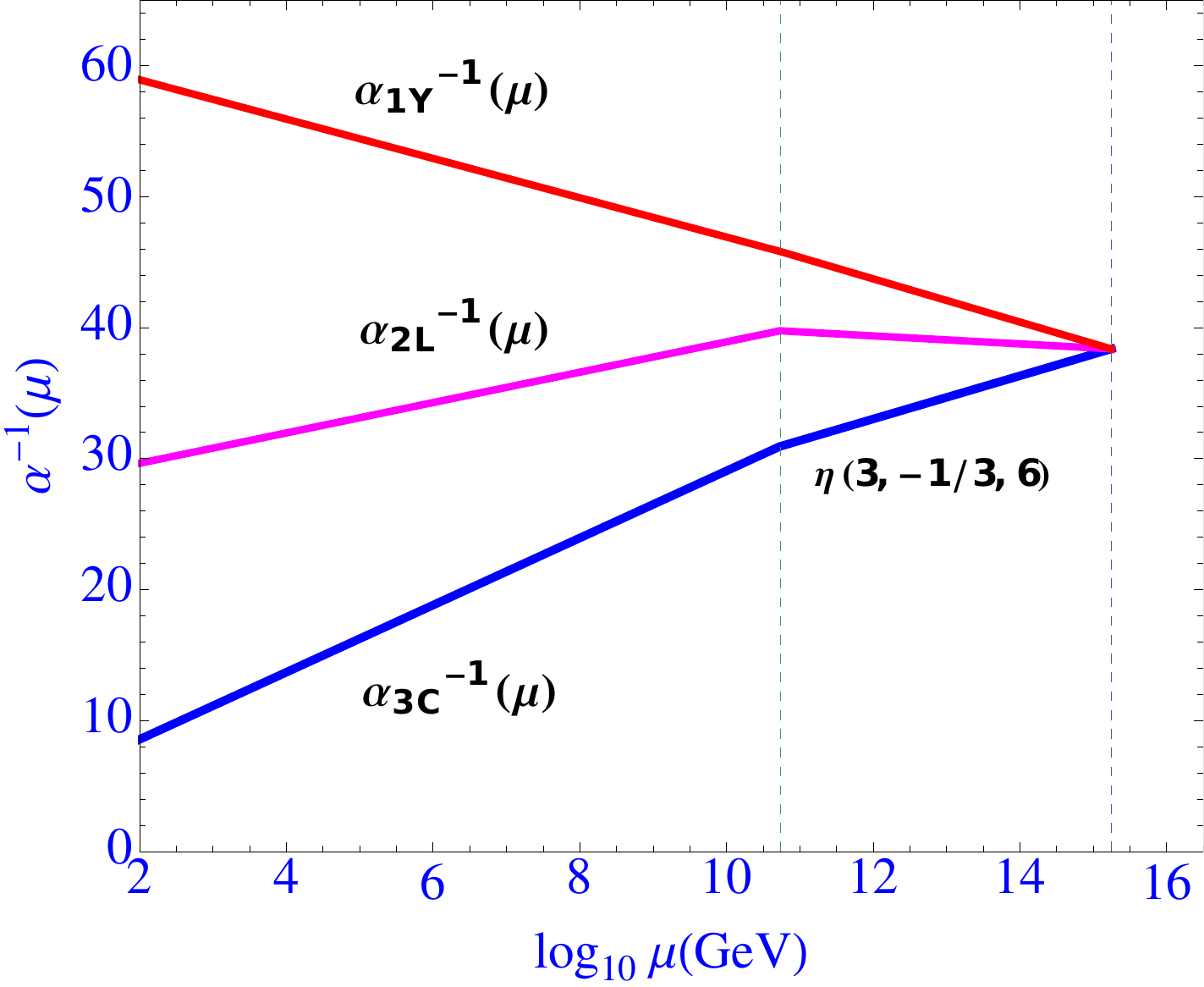}
\caption{Evolution of gauge couplings in Model-I with the real scalar submultiplet $\kappa
  (3,0,8)\subset {210}_H$ of
  mass  $M_{\kappa}=10^{9.2}$ GeV as depicted by the first vertical
  in the left panel. Unification of gauge couplings in Model-II due to the presence of the complex scalar component $\eta(3,-1/3,6)\subset {126}_H$ at $M_{\eta}=10^{10.7}$ GeV as shown in the right panel,}
 \label{fig:mod12}
\end{figure}

 In contrast to non-SUSY GUTs with one or more intermediate scales
it has been recently shown that unification is possible with only one
non-standard Higgs scalar $\kappa (3,0,8) \subset {210}_H$
\cite{Kynshi-mkp:1993,psc:2017,cps:2019} or $\eta (3,-1/3, 6) \subset
     {126}_H$ with the respective masses $M_{\kappa}=10^{9.2}$ GeV or
     $M_{\eta}=10^{10.7}$ GeV. Coupling unification is shown in Fig.\ref{fig:mod12}.
As outlined below these minimal SO(10) models have interesting applications in dark matter
decay \cite{psc:2017} manifesting as monochromatic PeV enegy neutrinos
detected at Ice-Cube \cite{IceCube}, Type-II seesaw prediction of
heavy scalar triplet leptogenesis in SO(10), verifiable LNV and LFV
decays with type-II seesaw neutrino mass \cite{mkp-rs:2018} in SU(5) along with WIMP DM
prediction, vacuum stability and observable proton decay \cite{psc:2017}.   
\subsection{Dark Matter Decay for PeV Energy IceCube Neutrinos}\label{sec:dd}
\begin{figure}[h!]
\includegraphics[scale=0.3]{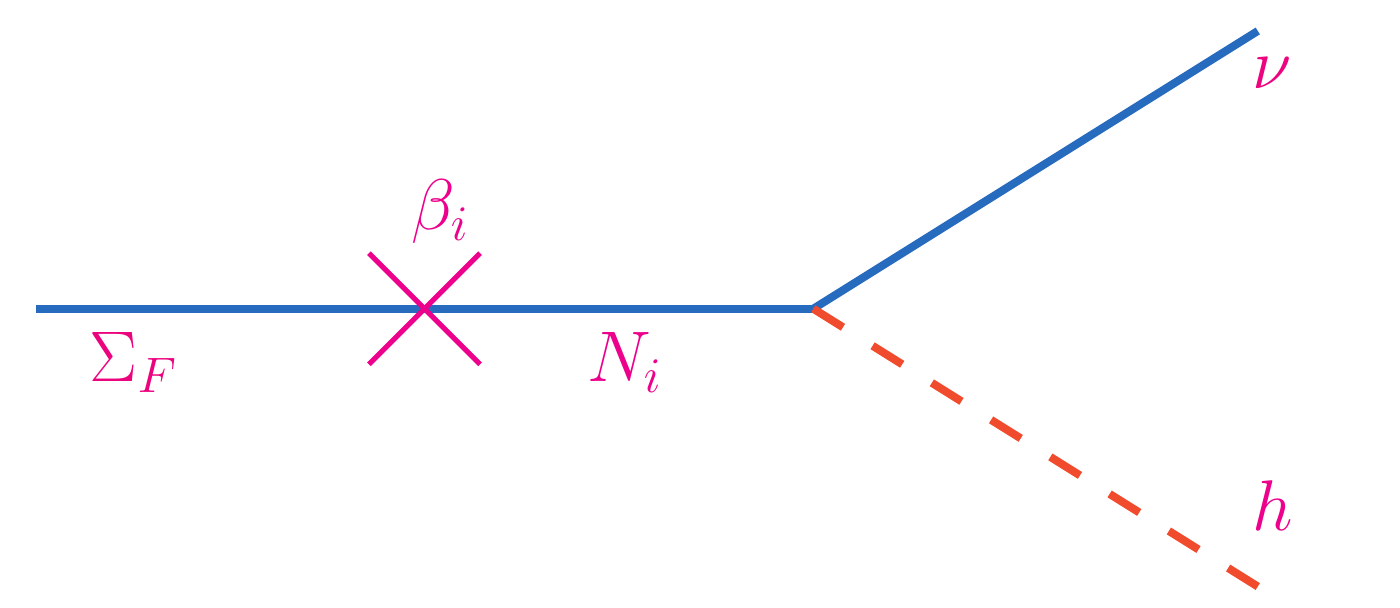}
\caption{Feynman diagram for dark matter decay $\Sigma_F\to \nu h$
  manifesting as monochromatic PeV energy neutrinos at IceCube.}
 \label{fig:dd}
\end{figure}

Each of the two models, Model-I and Model-II shown in Fig. \ref{fig:mod12} above,  predict
fermionic DM decay \cite{psc:2017} manifesting as monochromatic PeV energy neutrinos
detected recently at IceCube \cite{IceCube}. Both the models account for neutrino
mass via heavy RHN mediated canonical seesaw mechanism in concordance
with neutrino data. These models predict  three hierarchical RHNs which
 mix by different amounts with the fermionic singlet DM $\Sigma_F
(1,0,1) \subset {45}_F$ as a result of which the latter decays to
produce the standard Higgs and the PeV energy neutrino: $\Sigma_F\to
\nu h$. The decay mode is shown in Fig.\ref{fig:dd}
\subsection{Triplet Leptogenesis with New CP-Asymmetry
  Formulas} \label{sec:trl}
\begin{figure}[h!]
\includegraphics[scale=0.3]{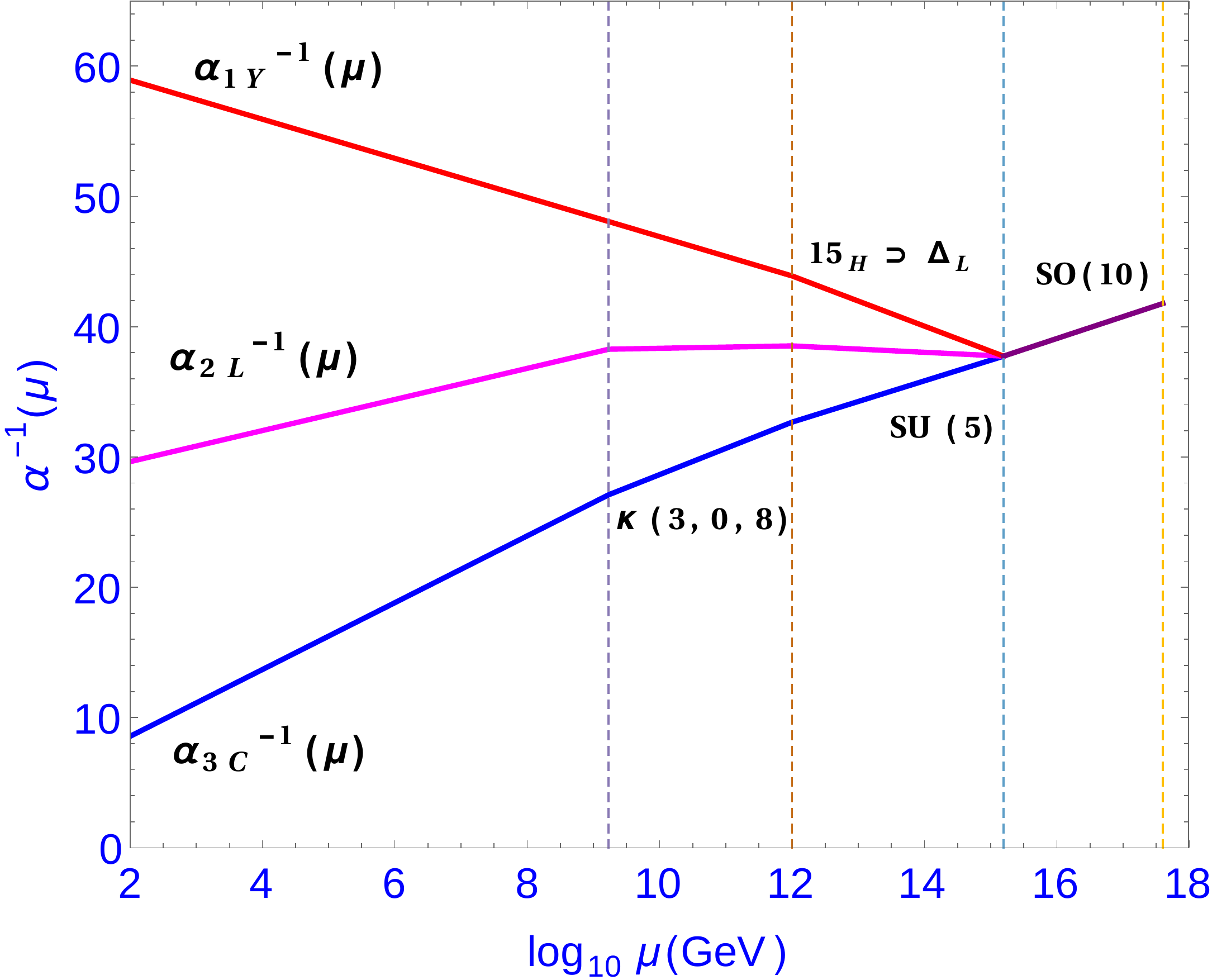}
\includegraphics[scale=0.3]{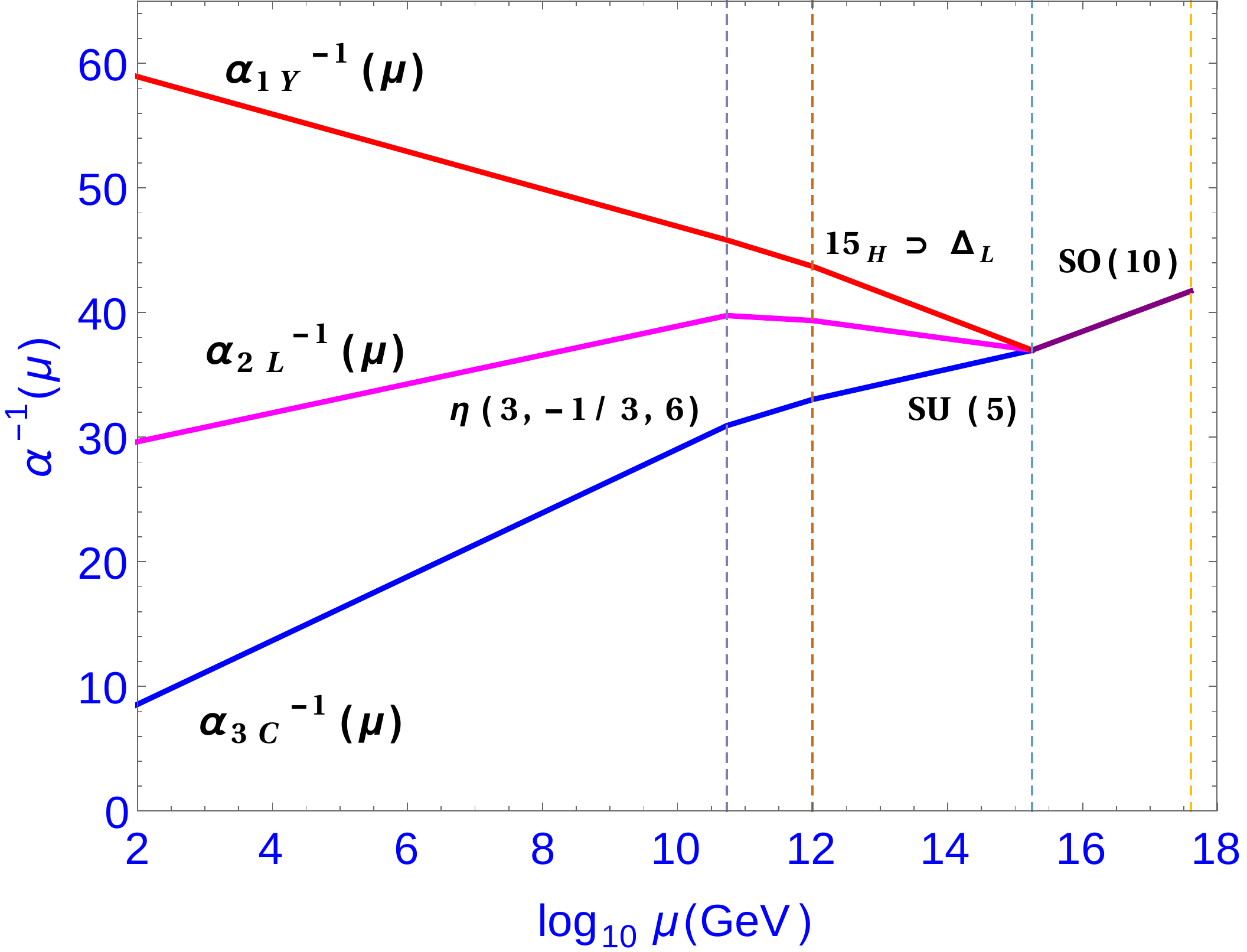}
\caption{Evolution of gauge couplings in Model-I with the real scalar submultiplet $\kappa
  (3,0,8)\subset {210}_H$ of
  mass  $M_{\kappa}=10^{9.2}$ GeV as depicted by the first vertical line
  in the left-panel. Unification of gauge couplings in Model-II due to the presence of the complex scalar component $\eta(3,-1/3,6)\subset {126}_H$ at $M_{\eta}=10^{10.7}$ GeV as shown in the right-panel,}
 \label{fig:kappaeta2}
\end{figure}

Ealier type-II seesaw dominance in SUSY or non-SUSY SO(10) was achieved
with an extended particle spectrum  near the TeV scale
\cite{mkp:2011,RNM-mkp:2011}. But as noted above, even without having such extended
spectrum near TeV scale, two minimal models \cite{Kynshi-mkp:1993,cps:2019} have been found to exhibit
type-II
seesaw dominance as they are also predicted by SO(10) breaking through
SU(5) route
\be
SO(10) \to SU(5)  \to SM. \label{eq:minchain}
\ee
The RG evolution of gauge couplings in the two corresponding models   
 are depicted through Fig. \ref{fig:kappaeta2} where 
in the left- panel (right-panel) unification is achieved  
 by $\kappa(3,0,8)$ ( $\eta(3,-1/3,6)$). 

Unlike such minimal grand desert modifications by only one non-standard lighter field below the GUT scale \cite{Kynshi-mkp:1993,psc:2017,cps:2019}, unification models also exist with more than one non-standard lighter fields \cite{Frig-Ham:2010,moreNS}. Whereas triplet fermionic DM $\sigma_F(3, 0,1)$ of mass $2.7$ TeV has been predicted in \cite{Frig-Ham:2010}, the three unification models discussed  in \cite{cps:2019} predict a real scalar siglet DM or a real scalar  singlet plus a fermionic triplet as DM with masses near $\simeq 1.0$ TeV. In addition they complete vacuum stability
of the scalar potential and predict baryon asymmetry through new CP-asymmetry formulas in triplet leptogenesis. Proton lifetimes predicted by all the three models are accessible to ongoing experimental searches. 
 \section{Summary and Outlook}
Besides the natural resolutions of gauge hierarchy problem and origin of three
forces of nature, SUSY GUTs also accomplish the desired expectations for dark
matter and their stability, and baryogenesis via leptogenesis while matching
the neutrino oscillation data through attractive seesaw
mechanisms. They can also predict LFV decays closer to current
experimental limits and verifiable proton decays . As pointed out SUSY SO(10) is capable of
representing all charged fermion masses with or without $S_4$ flavour
symmetry \cite{mkp:2008}. If neutrinos are quasi-degenerate, SUSY GUTs with
$S_4$ or $G_{224}\times S_4$, or even  $G_{2213}\times S_4$ unify quark and lepton mixings at high
scale and  are capable of answering the puzzle as to why neutrino mixings are so different.
It is high-time that  evidence of SUSY shows up at LHC energies \cite{LHCsusy,UC:2014,UC:2016}. Once the hitherto non-appearance of SUSY  is
reconciled with anthropic principles \cite{Weinberg,Barr}, a large
number of different GUT solutions with or without intermediate symmetries
are capable of resolving puzzles confronting the standard model
including gauge coupling unification, origins of
neutrino (and charged) fermion masses, baryon asymmetry of the
universe,  dark matter with matter parity \cite{Kadastik:2009}
as  stabilising gauged discrete symmetry, vacuum stability of
Higgs potential and 
 proton decay prediction accessible to ongoing experiments. A  novel ansatz for type-II seesaw dominance in a class of non-SUSY SO(10) not only predicts  new CP-asymmetry formulas for leptogenesis leading to baryon asymmetry of the universe, but it resolves the issues on
dark matter, vacuum stability, and verifiable proton decay
showing wide range of capabilities of these models  \cite{cps:2019}.
Two interesting unification possibilities through  minimal  grand desert modifications \cite{Kynshi-mkp:1993,cps:2019}
 by only one intermediate mass scalar  in each case and their various 
 applications in solving puzzles confronting the SM are emphasized.
\par\noindent {\bf Acknowledgment}\\
M.K.P. acknowledges financial assistance under the
  project SB/S2/HEP-011/2013 awarded by Govt. of India, New Delhi. R. S. acknowledges Ph.D. research fellowship by Siksha O Anusandhan, Deemed to be University.  
\par\noindent {\bf Author contribution statement}\\
M. K. Parida (Corresponding author) has made original contributions in different areas of High Energy Physics as cited and summarised in this paper which has been planned and outlined by him. Under the
supervision of M. K. P. the co-author Riyanka Samantaray has rederived 
all equations involved and understood the ideas pertaining to
 this work.  She has also carried out computations leading to different figures presented here.

\end{document}